\documentclass[aps,twocolumn,showpacs,amsmath,amssymb]{revtex4}
\usepackage{graphicx}% Include figure files
\usepackage{dcolumn}% Align table columns on decimal point
\usepackage{bm}% bold math
\usepackage{color}

\begin{document}

\title{Stochastic extension of the Lanczos method for nuclear shell-model 
  calculations with variational Monte Carlo method}
\author{Noritaka Shimizu$^1$}
\author{Takahiro Mizusaki$^2$}
\author{Kazunari Kaneko$^3$}

\affiliation{$^1$ Center for Nuclear Study, University of Tokyo, 
  Hongo, Tokyo 113-0033, Japan }
\affiliation{$^2$ Institute of Natural Sciences, Senshu University, Tokyo, 
  101-8425, Japan}
\affiliation{$^3$ Department of Physics, Kyushu Sangyo University, 
  Fukuoka 813-8503, Japan}

\date{\today}

\begin{abstract}
We propose a new variational Monte Carlo (VMC) approach 
based on the Krylov subspace
for large-scale shell-model calculations. 
% which can overcome the exponential growth of 
% the dimension of the Hilbert space
% in the conventional Lanczos diagonalization method.
A random walker in the VMC is formulated with 
the $M$-scheme representation, and samples 
a small number of configurations from a whole Hilbert space stochastically.
This VMC framework is demonstrated in the shell-model 
calculations of $^{48}$Cr and $^{60}$Zn, 
and we discuss its relation to 
a small number of Lanczos iterations.
% This approach provides us with a systematic improvement of the 
% wave function with conventional particle-hole-excitation truncation.
By utilizing the wave function obtained by 
the conventional particle-hole-excitation truncation as an initial state, 
this VMC approach provides us with a sequence of 
systematically improved results. 
\end{abstract}


\maketitle

\section{Introduction}
\label{sec:intro}

Large-scale shell model calculations 
in the nuclear structure study are performed by 
solving an eigenvalue problem of a large-dimension sparse matrix. 
The Krylov-subspace method is one of the best tools, 
and practically the only solution to solve the problem efficiently \cite{krylov}.  
The Krylov subspace is spanned by a starting vector $v$ and 
the product of the first $p$ powers of the Hamiltonian matrix $H$, 
namely $\{ v, Hv, H^2v, \cdot \cdot \cdot,  H^pv \}$.
An eigenvalue of the Hamiltonian matrix in Krylov subspace is called 
the Ritz value, and it is known to converge in a small number of iterations
to the largest (or smallest) eigenvalue, 
since $H^p v$ with large $p$ is dominated by the eigenvectors 
which have the large absolute eigenvalue.
In the Krylov-subspace method, the converged Ritz value 
becomes a good approximation to the exact eigenvalue.
Since the construction of the Krylov subspace 
requires only the matrix-vector product, the Krylov-subspace method 
has been extensively used 
and developed to solve an eigenvalue problem 
of a huge sparse matrix.

In the case of nuclear shell-model calculations, 
the Hamiltonian matrix in $M$-scheme basis is very sparse 
since the Hamiltonian consists of one-body and two-body interactions.
The needed dimension  of the Hamiltonian matrix is often 
quite huge, therefore, the Krylov-subspace 
iteration algorithm is quite efficient.
The Lanczos algorithm, one of the most famous Krylov subspace algorithms, 
was introduced in 70's, 
\cite{lanczos, lanczos-sebe, M-lanczos}
and has been widely used in shell-model calculations. 
Nowadays, it is implemented to take advantage of 
massively parallel computations \cite{mfdn}.

Nevertheless, the application is still hampered by the exponential 
growth of the dimension of the Hilbert space. 
The size of a state vector surpasses the capacity of memory
and the truncation of the model space is required.
The most naive truncation is to assume a Fermi level 
for single-particle occupation and to restrict the number of 
particle-hole excitation across the Fermi level up to $t$ particles.
It is called the $t$-particle $t$-hole truncation 
and frequently used in practical calculations.
As $t$ increases, the eigenstate in the truncated subspace 
approaches the true eigenstate rather gradually.
However, the significance of the large $t$ component remains
and is difficult to estimate (e.g. \cite{horoi_ni56}) 
due to the limitation of computational resources.

On the other hand,  much effort has been paid to circumvent 
this problem by introducing the Markov Chain Monte Carlo (MCMC)
in the context of quantum Monte Carlo
\cite{smmc,rombouts}. However, the notorious 
``sign problem'' prohibits to use realistic shell-model interaction 
practically.

In the present work, to estimate the omitted contribution of 
$t$-particle $t$-hole truncation for 
the exact shell-model energy in full Hilbert space
we introduce the variational Monte Carlo (VMC) approach into 
the shell-model calculations, and 
show that this VMC can overcome the limitation of the truncation scheme 
without treating the basis vectors of the full Hilbert space 
explicitly.
% We have introduced the VMC in Ref. \cite{vmcsm}, 
% however

% This VMC approach is formulated by a random walker in $M$-scheme 
% in Sect.\ref{sec:mcmc}.
% The determination of the variational parameters 
% utilizing the reweighting method is described in Sect.\ref{sec:reweight}.
% Sect.\ref{sec:discuss} shows the numerical results and their discussions.

% While we have to 
% treat directly the huge vectors stored on memory in the Lanczos diagonalization method.

\section{Formulation}
\label{sec:formulation}

We describe the form of a trial wave function of this 
VMC approach.
At the beginning, we calculate the lowest eigenstate
in the $t$-particle $t$-hole truncated subspace, $|\psi^t\rangle$. 
Then, we introduced the VMC approach 
combined with the Krylov-subspace method with $|\psi^t\rangle$
being a starting vector 
to improve the approximation systematically.

The Ritz vector, or the approximated eigenvector on the Krylov subspace 
is taken as a trial wave function such as 
\begin{equation}
  |\Psi({\bf c}) \rangle 
  = \left( \sum_{q=0}^{p} c_q H^q \right)| \psi^t \rangle
  \label{eq:wf}
\end{equation}
where $H$ is the Hamiltonian and 
${\bf c} = \{c_0, c_1,c_2,...,c_p\}$ is 
a set of variational parameters, which are 
determined by minimizing the energy expectation value.
% $E({\bf c}) = \langle \Psi({\bf c})| H |\Psi({\bf c}) \rangle 
% / \langle \Psi({\bf c})|\Psi({\bf c}) \rangle $.
The $|\psi^t\rangle$ is represented by 
a linear combination of the 
$M$-scheme basis states, that is, 
$|m\rangle = c^\dagger_{m_1}c^\dagger_{m_2}...c^\dagger_{m_A}|-\rangle$
where $c^\dagger_{m}$ is a creation operator of the single-particle state, $m$, 
and $|-\rangle$ stands for an inert core.

The energy expectation value of the trial wave function 
is written by inserting the complete set of the $M$-scheme subspace, 
$1 = \sum_{m \in \{M^\pi\}} |m\rangle \langle m|$, 
where $|m\rangle$ has a fixed  $z$-component of angular momentum $M$ and parity $\pi$, 
such as 
\begin{eqnarray}
  \label{eq:eval}
  E({\bf c})
  &=&  \frac{\langle \Psi| H |\Psi\rangle}
  {\langle \Psi |\Psi \rangle}% \nonumber \\ &=&
  =  \frac{ \sum_{m \in \{M^\pi\}} |\langle m | \Psi \rangle|^2 
    \frac{\langle m | H |\Psi \rangle}{\langle m | \Psi \rangle} }
  {\sum_{m \in \{M^\pi\}} |\langle m | \Psi \rangle|^2} .
\end{eqnarray}
Since the number of $M$-scheme states rapidly 
increase as a function of the particle number and 
the size of the single-particle space, 
the practical summation of $|m\rangle$ in Eq.(\ref{eq:eval}) 
often becomes difficult to perform. 
Therefore, we apply the Monte Carlo technique to this summation 
and calculate the energy expectation value $E({\bf c})$ 
stochastically. 
The Monte Carlo random walker of $m$ state is formulated 
in the $M$-scheme basis obeying 
the probability proportional to $|\langle m |\Psi({\bf c})\rangle|^2$.
The complete set of $|m\rangle$ is represented 
by the $N$ samples of the random walkers as
\begin{eqnarray}
  E({\bf c})
  &\sim& 
  %\frac{1}{N} \sum_{m \in {\rm M.C.} }
  %\frac{\langle m | H |\Psi ({\bf c})\rangle}{\langle m | \Psi ({\bf c})\rangle}
  % \nonumber \\
  % &=& 
  \frac{1}{N} \sum_{m \in {\rm M.C.} } E_{\rm local}(m, {\bf c})
  \label{eq:ev}
\end{eqnarray}
with the local energy 
\begin{equation}
  E_{\rm local}(m, {\bf c}) 
  =    \frac{\langle m | H |\Psi ({\bf c})\rangle}{\langle m | \Psi ({\bf c})\rangle}
\end{equation}
and the sampling density 
\begin{equation}
  \rho_{\bf c}(m) = 
  \frac{|\langle m |\Psi ({\bf c})\rangle|^2}{
    |\sum_{m'} \langle m' | \Psi ({\bf c})\rangle|^2} .
  \label{eq:rho} 
\end{equation}
The $\sum_{m \in {\rm M.C.} } $ denotes 
the summation of $N$ samples, $m$, 
which are generated by 
the $M$-scheme random walker with the probability density $\rho_{\bf c}(m)$.

In order to compute the $E({\bf c})$ in Eq.(\ref{eq:ev}) stochastically, 
we briefly describe the MCMC process 
to generate a random walker in $M$-scheme basis states.
It was first introduced in Ref. \cite{vmcsm}.

The transition of the random walker is controlled by the Metropolis algorithm.
The candidate of the transition is created by 
two-particle two-hole operator like 
$|m'\rangle = c^{\dagger}_ic^{\dagger}_jc_kc_l|m\rangle$ 
with $c^\dagger_i$ being the creation operator of a single-particle
state $i$.
The indices, $i,j,k,$ and $l$, are restricted 
so that $|m\rangle$ and $|m'\rangle$ having the same 
$z$-component of angular momentum and parity.
Whether or not a random walker $|m\rangle$ moves to  $|m'\rangle$ 
depends on the ratio $p_{\bf c}(m') = \rho(m')/ \rho(m)$ as 
\begin{equation}
  p_{\bf c}(m')=\left|
    \frac{\langle m'|\Psi({\bf c}) \rangle }{\langle m|\Psi({\bf c}) \rangle}
  \right|^{2}.
\end{equation}
If $p_{\bf c}(m')\geq 1$, the walker $|m\rangle$ always moves to $|m'\rangle$. 
If $p_{\bf c}(m')<1$, according to the $p_{\bf c}(m')$, we determine whether or not the walker $|m\rangle$ moves 
to $|m'\rangle$.
This procedure satisfies the detailed balance condition and ergodicity.

The $M$-scheme walker is automatically restricted 
to good $M$, parity, $z$-component of isospin subspace 
because the initial wave function $|\psi^t\rangle$ is already 
an eigenstate of these quantum numbers, and the sampling density 
not having suitable values of $M$ and parity becomes zero. 
In addition, because the $|\psi^t\rangle$ is an eigenstate of 
the total angular momentum $J^2$
and the Hamiltonian commutes with the $J^2$,  
the expectation value of $J^2$ of the resultant state in Eq.(\ref{eq:wf}) is 
the same as that of the $|\psi^t\rangle$ 
without statistical error. Therefore,  unlike 
the formulation of the VMC in Ref.\cite{vmcsm}, 
we do not need angular-momentum projection 
which is represented by three-dimensional integral 
over the Euler angles
and causes a time-consuming numerical computation.

In practical calculations, 
we implemented a cache algorithm for efficient computation. 
Namely, we store any calculated matrix elements 
$\langle m |H^q|\psi^t\rangle$ with $0\leq q \leq p$ in memory, and 
reuse them if needed. This prescription shortens 
the computational time drastically, 
however it requires a large amount of memory in compensation.
This is a trade off between the computational time and the memory usage, 
and the introduction of an efficient cache algorithm would ease this problem.

Here, we discuss the relation with the power method \cite{powerlanczos, genlanczos}, 
which has a simple form of the trial wave function and 
has been used widely.
In the power method, the wave function is approximated by 
\begin{equation}
  |\Psi(\Lambda) \rangle =  (\Lambda - H)^p |\psi^t \rangle , 
  \label{eq:power}
\end{equation}
with a constant value $\Lambda$.
This wave function is equal to Eq.(\ref{eq:wf}) if {\bf c} is 
taken as a set of binomial coefficients.
Since the VMC has a larger variational space 
than the power method with the same $p$, 
the VMC provides a better approximation. 
Equation(\ref{eq:power}) is extended to 
$ \prod_{q=1}^p (\Lambda_q - H) |\psi^t \rangle $, 
which is equivalent to Eq.(\ref{eq:wf}).
The optimization of the variational parameters 
requires negligible computational cost 
compared to that of calculation without variation by utilizing 
the reweighting method discussed below.

For the energy minimization with respect to the variational parameters 
${\bf c}$,  
the Nelder-Mead downhill simplex method is utilized \cite{num_recipe}.
In this method, the energy functional is minimized by 
morphing the polytope consisting of $p+1$ vertexes 
in the $p$-dimension parameter space.
No other information of the energy functional is required, 
e.g. energy gradient.
Thus we have to obtain the energy 
expectation values with many sets of samples with various ${\bf c}$'s. 
We describe the reweighting method \cite{reweighting} to determine 
the best variational 
parameters with a single set of samples for a certain set of ${\bf c}$. 

We suppose that a set of Monte Carlo samples and $E({\bf c_0})$ has already 
been obtained with appropriate initial parameters ${\bf c_0}$.
Then, we calculate the energy expectation value 
$E({\bf c'})$ 
where ${\bf c'}$ is close enough to ${\bf c_0}$. 
The $E({\bf c'})$ is written as
\begin{eqnarray}
  \label{eq:reweight}
  E({\bf c'})
  &=&  
%  \frac{ \sum_{m \in \{M^\pi\}} |\langle m | \Psi({\bf c'}) \rangle|^2 
%    \frac{\langle m | H |\Psi({\bf c'}) \rangle}{\langle m | \Psi({\bf c'}) \rangle} }
%  {\sum_{m \in \{M^\pi\}}  |\langle m | \Psi({\bf c'}) \rangle|^2 .
%  } 
  \frac{ \sum_{m \in \{M^\pi\}} |\langle m | \Psi({\bf c_0}) \rangle|^2 
    \left| \frac{ \langle m | \Psi ({\bf c'})\rangle }{\langle m | \Psi({\bf c_0}) \rangle} \right|^2 
    \frac{\langle m | H |\Psi({\bf c'}) \rangle}{\langle m | \Psi({\bf c'}) \rangle} }
  {\sum_{m \in \{M^\pi\}}  |\langle m | \Psi({\bf c_0}) \rangle|^2
    \left| \frac{ \langle m | \Psi({\bf c'}) \rangle }{\langle m | \Psi({\bf c_0}) \rangle} \right|^2 .
  } 
\end{eqnarray}
By using the random walker of $\rho_{\bf c_0}(m)$, 
it is evaluated stochastically using
\begin{eqnarray}
  E({\bf c'})
  &\sim & \frac{ \frac1N \sum_{m \in {\rm M.C.} } 
    R(m, {\bf c', c_0})  E_{\rm local}(m, {\bf c'}) }
  { \frac1N \sum_{m \in {\rm M.C.} } R(m, {\bf c', c_0}) }
\end{eqnarray}
with the reweighting factor 
\begin{equation}
  R(m, {\bf c', c_0}) = 
  \left| \frac{ \langle m | \Psi ({\bf c'})\rangle }{\langle m | \Psi({\bf c_0}) \rangle} \right|^2 .
\end{equation}
Note that $\sum_{m \in {\rm M.C.}} $ in Eq.(\ref{eq:reweight}) 
denotes the Monte Carlo summation of 
the $M$-scheme random walkers, of which 
the sampling density is $\rho_{\bf c_0}(m)$, 
not $\rho_{\bf c'}(m)$.
In practice, when we evaluate the $E({\bf c_0})$
we generate a set of samples with the density distribution $\rho_{\bf c'}(m)$
and keep all the computed matrix elements,  
$\langle m |H^q|\psi^t\rangle$. 
By reusing this random walker and the matrix elements, 
we do not need additional computations to compute the $E({\bf c'})$.
% We do not need additional computations at the reweighting process to try 
% different ${\bf c'}$, 
% by storing all the matrix elements 
% $\langle m |H^q|\Psi\rangle$ once calculated. 

\section{Numerical results}
\label{sec:results}

\begin{figure}[htbp]
  \includegraphics[scale=0.30]{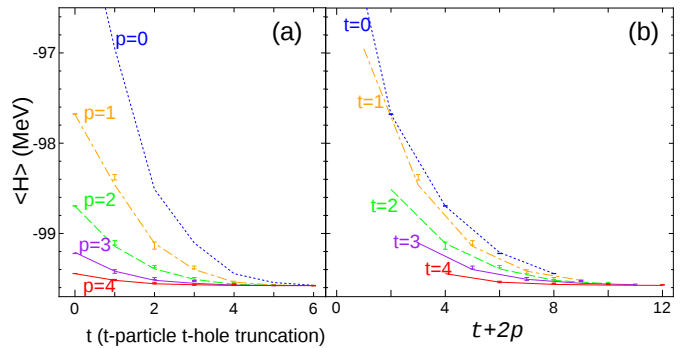}
  \caption{(Color online) Energy expectation values of
    the $J^\pi=0^+$ trial wave functions of $^{48}$Cr in $pf$ shell.
    (a) the blue, green, purple, and red
    error bars denote the VMC results with $p=0,1,2,3,$ and $4$, respectively.
    (b) the blue, green, purple, and red error bars denote 
    the VMC results with $t=0, 1, 2, 3,$ and $4$ 
    as functions of $t+2p$.
    The lines denote the results of the corresponding 
    $p$-iteration Lanczos method.
  }
  \label{cr48j0}
\end{figure}

To present the feasibility of the present method, we demonstrate 
the shell-model calculation of $^{48}$Cr in $pf$ shell as an example.
Figure \ref{cr48j0} shows the energy expectation values of the ground-state energy 
of the $^{48}$Cr with GXPF1A interaction \cite{gxpf1a}.
Its $M$-scheme dimension is 1,963,461. 
We generate 48 random walkers utilizing the Metropolis algorithm. 
For each walker, after we run 
1000 steps as burn-in process, 
a random walker moves 10000 steps in the $M$-scheme space.
The variational parameters $\bf c$ in Eq.(\ref{eq:wf}) 
were determined to minimize the energy by the Nelder-Mead method 
and the reweighting technique discussed 
as before.

Figure \ref{cr48j0} also  shows 
the Ritz value of the Lanczos method with $p$-step iterations  
and $|\psi^t\rangle $
being the initial vector. 
The $p$ and $t$ in Fig.\ref{cr48j0}(a)
denote the power $p$ and the number of particle-hole truncation 
of the wave function $|\psi^t\rangle$ in Eq.(\ref{eq:wf}).
The exact shell-model energy is $-99.578$MeV, which is 
almost reached at $p=3,4$ and  $t\geq 3$.
If the variational parameters of the VMC were determined without statistical error, 
this trial wave function would equal the wave function 
obtained by the Lanczos method with $p$-iterations.
More generally, this is a Monte Carlo formulation of Krylov subspace 
technique because the trial wave function is a linear combination  of 
$H^q |\psi\rangle $ terms with $0\leq q \leq N$.
The VMC results (error bars) agree quite well with those of Lanczos method 
(lines), which means that the Nelder-Mead optimization with reweighting 
works quite well. 
While the energy of the initial state $|\psi^t\rangle$, or $p=0$ error bars 
in Fig.\ref{cr48j0}(a),  
does not converge well as a function of $t$, 
the VMC values of $p=3,4$ converges well even for the small $t$.

Since the Hamiltonian only contains two-body interactions, 
$H^p|\psi^t \rangle $ is considered to 
be a state in $(t+2p)$-particle  $(t+2p)$-hole truncated subspace.
Figure \ref{cr48j0}(b) shows the 
energy as a function of $t+2p$. 
The left ends of these lines on Fig.\ref{cr48j0}(b) 
correspond to the exact solution with $t$-particle $t$-hole truncation, 
and therefore the variational lower bound of the truncated subspace.
With increasing $p$, the energy converges well and 
close to the Lanczos value.

\begin{figure}[htbp]
  \includegraphics[scale=0.32]{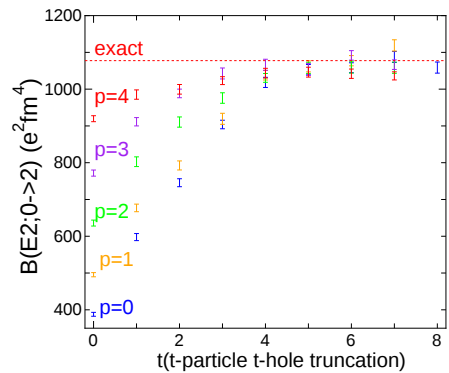}
  \caption{(Color online) B(E2;$0^+\rightarrow 2^+$) transition probabilities
    of $^{48}$Cr provided by the VMC calculations against truncation scheme $t$.
    The ``exact'' value is obtained by  $t=8$ Lanczos shell-model calculation.
  }
  \label{cr48be2}
\end{figure}
Physical observables other than the energy, e.g.  moments and 
transition probabilities, are also obtained 
in a similar manner to Ref.\cite{vmcsm}.
Figure \ref{cr48be2} shows the B($E2$) transition probability
obtained by the VMC. 
The effective charge is $(e_p,e_n)=(1.2,0.8)e$.
The convergence behaves similar to the case of the energy 
in Fig.\ref{cr48j0}(a)
and the VMC dramatically improve the $E2$ values 
in the region of small $t$.

For the demonstration 
of larger-scale calculations, we discuss the case of 
$^{60}$Zn in $pf$ shell with the GXPF1A interaction \cite{gxpf1a}.
The $M$-scheme dimension is huge, $2.2\times 10^9$, 
though it is somehow tractable by the recent shell-model code \cite{mshell64}.
\begin{figure}[htbp]
  \includegraphics[scale=0.3]{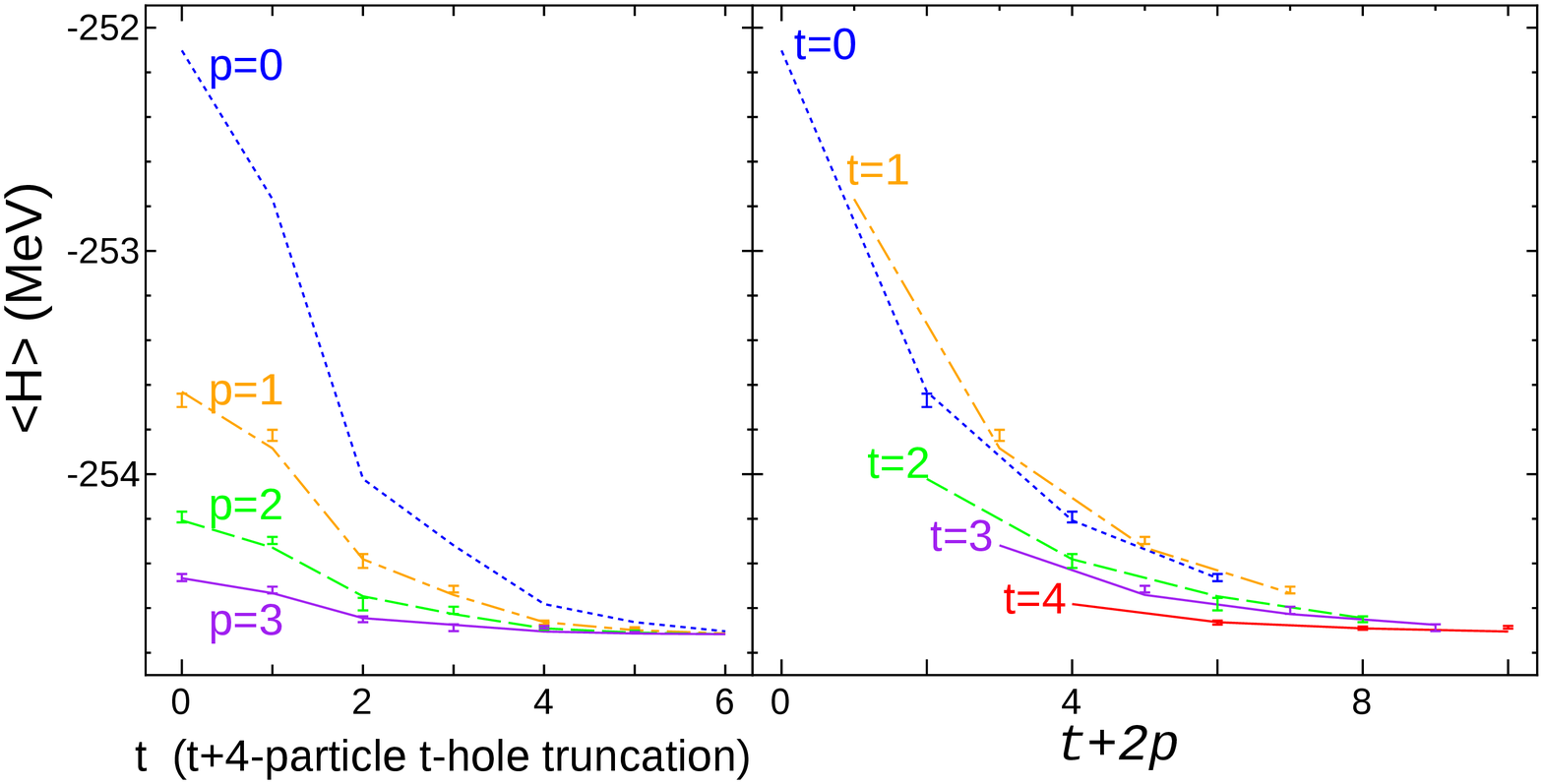}
  \caption{(Color online) Energy expectation values of
    the $J^\pi=0^+$ trial wave functions of $^{60}$Zn in $pf$ shell.
    See the caption of Fig.\ref{cr48j0}.
  }
  \label{zn60_j0}
\end{figure}

Figure \ref{zn60_j0} shows the energy expectation values 
of the ground state  obtained by the VMC and 
the corresponding Lanczos method. 
Though the dimension is about $10^3$ times larger than that of $^{48}$Cr, 
the energy convergence is similar to Fig.\ref{cr48j0}. 

\begin{figure}[htbp]
  \includegraphics[scale=0.32]{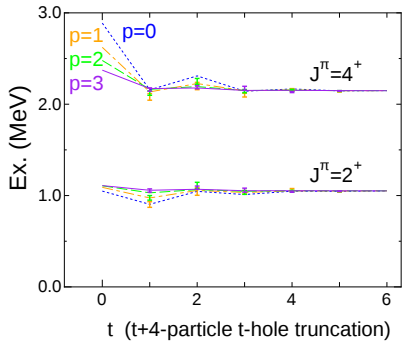}
  \caption{(Color online) Excitation energies of
    $J^\pi=2^+$ and $4^+$ states  of $^{60}$Zn in $pf$ shell.
    See the caption of Fig.\ref{cr48j0}.
  }
  \label{zn60_ex}
\end{figure}

Figure  \ref{zn60_ex} shows the excitation energies 
obtained by the VMC approach and corresponding $p$-iterated Lanczos method.
The excitation energies converged far faster than the energy itself, 
and the VMC significantly improves the convergence. 

\section{Summary}
\label{sec:summary}

In summary we proposed a new VMC formalism to improve the conventional 
particle-hole truncation shell-model calculations systematically, 
and demonstrated its feasibility at $^{60}$Zn in $pf$ shell. 
In this formalism, we do not have the sign problem because it is implemented 
by the variational Monte Carlo, namely the density probability 
is defined by the wave function squared.
However, the power of the Hamiltonian in the 
Monte Carlo implementation of the Lanczos method is 
restricted to be rather small, such as $p=3$ or $4$. Nevertheless, 
it provides us with good convergence of the energy 
and is expected to be useful to go beyond the conventional 
diagonalization method. 
The energy expectation value of our $p$-power VMC scheme agrees with
that of $p$-th step Lanczos 
method in full space within statistical error. 
It means that the MCMC procedure and reweighting method 
to determine the variational parameters work well and stable.
The present VMC approach
provides us with the facile estimation of the exact energy eigenvalue 
by using the $t$-particle $t$-hole truncated wave functions.
The energy variance can also be calculated stochastically 
with the same formulation, which is expected 
to be helpful for the energy-variance extrapolation technique
\cite{variance_extrap}, and it remains for future study. 

The cache algorithm to store $\langle m |H^q|\phi^t\rangle$ with 
a random walker $m$  drastically reduces the computation time 
At present, since we store all matrix elements on memory, 
it requires a large amount of memory usage. 
This difficulty can be eased by sophisticated cache algorithm, 
the implementation of which is important for further applications.

\section*{Acknowledgement}
\label{sec:acknowledgement}

We are grateful to Dr. J. Anderson for 
carefully reading the manuscript. 
This work has been supported by 
the SPIRE Field 5 from MEXT, 
the CNS-RIKEN joint project
for large-scale nuclear structure calculations, 
and Grants-in-Aid 
for Scientific Research (No. 20201389) from JSPS, 
Japan.
The Lanczos shell-model calculations were performed 
by the code MSHELL64 \cite{mshell64}.

%%%%%%%%%%%%%%%%%%%%%%%%%%%%%%%%%%%%%%%%%%%%%%%%%%%%%%%%%%%%%%%%%%%

\end{document}